# Direct observation of local K variation and its correlation to electronic inhomogeneity in (Ba1-xKx)Fe2As2 Pnictide

W. K. Yeoh[1,*], B. Gault[1], X. Y. Cui[1], C. Zhu[1], M. P. Moody[1], L. Li[1], R. K. Zheng[1,*], W. X Li[2], X. L. Wang[2], S. X. Dou[2], C. T. Lin[3] and S. P. Ringer[1]

[1]Australian Centre for Microscopy & Microanalysis, University of Sydney, NSW 2006, Australia
[2]Institute for Superconducting & Electronic Materials, University of Wollongong, NSW 2519, Australia
[3] Max-Planck-Institut für Festkörperforschung, Heisenbergstr. 1, D-70569 Stuttgart, Germany

Local fluctuations in the distribution of dopant atoms are a suspected cause of nanoscale electronic disorder or phase separation observed within the pnictide superconductors. Atom probe tomography results present the first direct observations of dopant nano-clustering in a K-doped 122-phase pnictides. First-principles calculations suggest the coexistence of static magnetism and superconductivity on a lattice parameter length scale over a large range of doping concentrations. Collectively, our results provide evidence for a mixed scenario of phase coexistence and phase separation originating from variation of dopant atom experiments distroibutions.

Since the discovery of the superconductivity in iron pnictide [1-3], the phase diagram has been a focus of intense research and debate due to the competition between the antiferromagnetic spin-density-wave (AFM-SDW) and the superconducting (SC) states. There is a growing evidence indicating that magnetically ordered phases and SC states are mesoscopically/microscopically separated [4-7] or even coexist [8-11] in the K or Co doped BaFe2As2 (122-FeAs) system. Most of the studies on K or the Co doped samples to date are in favor for the coexistence of magnetic order and superconductivity [8-11] and have consistently ruled out the presence of phase separation [12-14]. Conversely, Park *et al.* [4] have proposed electronic phase separation of the AFM and SC and the model has been consequently supported by others studies via the muon spin-relaxation [5], $^{75}$As NMR [6] and neutron-diffraction measurements [7].

As the magnetism and SC evolve with the doping level, the coexistence or phase separation is likely to be correlated to local dopant distribution. Nevertheless, lack of direct experimental evidence at the atomic scale means this mystery remains unsolved. Atom probe tomography (APT) [15] provides three-dimensional chemical mapping with near-atomic resolution and therefore is ideally suited to investigate the distribution of dopants and rationalize the origin of the electronic disorder in 122-FeAs-based superconductors. *To the authors' knowledge, the present findings constitute the first direct observation of inhomogeneous distribution of K at the atomic scale.* We demonstrate that non-uniform distributions of dopants induce broad $T_c$ distribution, where AFM and SC may coexist on the lattice parameter length scale.

(Ba0.72K0.28)Fe2As2 crystals were obtained by the self-flux method detailed in ref. [16]. A superconducting transition temperature $T_c$ occurred at 32 K for the doped Ba0.75K0.25Fe2As2 [17] similar to the value in ref. [18]. The

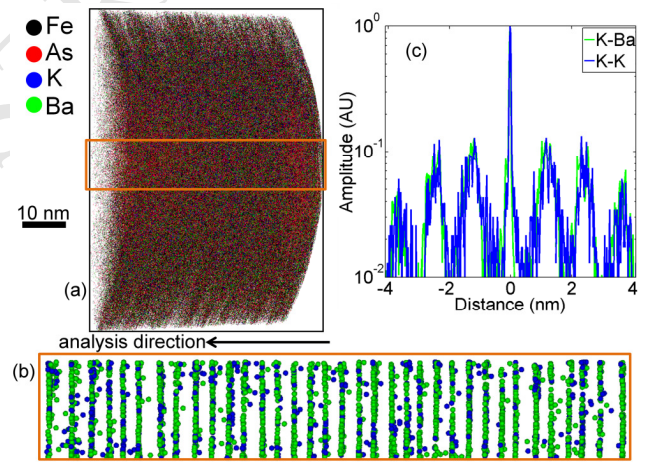

Fig. 1 (a) 3D APT reconstruction of (Ba0.72K0.28)Fe2As2, superconductor. Each dot represents individual atom of K, Ba, As and Fe (b) Enlarged of sub-volume from (a) shows lattice planes of K and Ba. Distance between each planes is $c$/2 (c) 1D spatial distribution maps in the analysis-direction highlights the distribution of K and Ba in atomic planes.

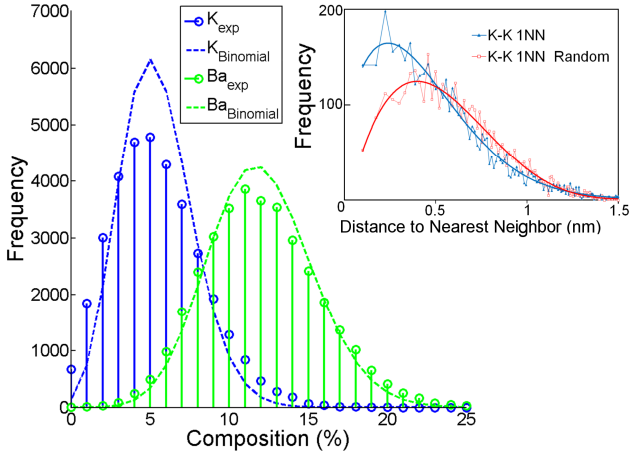

Fig. 2 . (a) Frequency distribution of Ba and K atoms. Dashed line shows the binomial distribution expected for a random distribution. (b) Nearest neighbor distribution of K for the experimental data and also the randomized case. Solid line is the fitting result for both distributions.

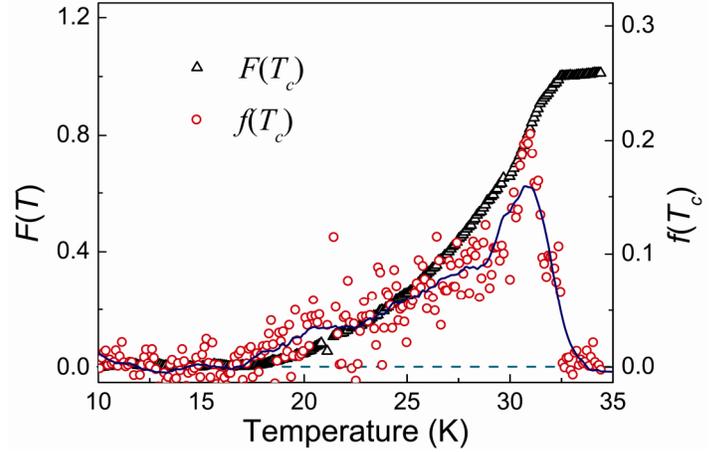

Fig. 3 $T_c$ distribution in K doped BaFe$_2$As$_2$ single crystals, obtained by deconvolution of the calorimetric data. Broader $T_c$ distribution is due to the inhomogeneous K distribution. the data.

specific heat was measured from 1 to 35 K at zero field and 12 T, using a long relaxation technique, described elsewhere [19]. Needle-shaped atom probe specimens with tip radius of < 100 nm were prepared by focused ion beam (FIB) using the lift-out method described in ref. [20] in dual beam instruments (FEI Quanta 200 and xT Nova NanoLab 200). Specimens were annular milled and cleaned with low energy ion milling to limit the damage induced by Ga implantation. Pulsed-laser APT was performed using a LEAP 3000X Si (Cameca Corp.). APT generates a highly chemically and spatially resolved 3D atomic map of the specimen [21]. Atom probe analyses were performed with a flight length of 90 mm, at a base pressure of 3 x 10$^{-11}$ torr, a temperature of 20 K with a laser pulse energy of 0.2 nJ, at a wavelength of 532 nm, focused on a spot of diameter < 10 □m and at a repetition rate of 200 kHz.

Fig. 1 (a) presents an APT reconstruction of K doped BaFe$_2$As$_2$ superconductor, generated from a data set of > 10$^6$ atoms with the Fe, As, Ba and K atoms displayed. Single atoms are represented by dots, and the dot color indicates its chemical identity as determined by the time-of-flight spectrometry. The measured Ba : K : Fe : As ratio within the reconstruction was 1.92: 0.68 : 3.7 : 3.7, close to the expected 1.5 : 0.5 : 4 : 4. It is noteworthy that the 1:1 ratio of Fe and As and 3:1 for Ba and K are preserved. Structural information can be obtained by means of spatial distribution maps (SDM) [22] in the in-depth direction. This is effectively a 1D radial distribution function in the reconstructed (001) direction. A spatial distribution histogram of Ba and K atoms relative to Fe atoms in the $c$-axis is plotted in Fig1(c). Three obvious peaks corresponding to interspacing value of 1.15 nm, 2.2nm and

3.8 nm can be detected within 4 nm distance with Fe planes as the reference planes. This value agrees closely with the expected value of interspacing in this direction between Fe and its nearest Ba/K neighbors. The SDM analysis also confirms that B and K reside in the same plane due to the overlapped of normalized intensity signal in the same position that matching to the $c$/2 distance.

To determine the nanoscale distribution of K and Ba a frequency distribution analysis was undertaken, whereby, the data set was divided spatially into blocks of 100 atoms, and the K and Ba compositions within each block were determined, as shown in Figure 2a. The histogram peaks indicate that K and Ba have the average composition of 0.12 and 0.05 respectively which are close to the nominal value. Each frequency distribution was compared directly to the corresponding binomial distribution, i.e. the distribution expected if the solute was distributed randomly throughout The K frequency distribution was found to have considerable deviation from the binomial distribution with $\chi^2$ test value ($\chi^2$ = 629.24 for 15 degrees of freedom) implying significant non-uniform distribution. The Ba frequency distribution was more consistent with the expected binominal distribution, with $\chi^2$ test value ($\chi^2$ = 69.53 for 23 degrees of freedom) indicating a more homogeneous distribution compared to K atoms. Nearest neighbor (NN) analyses [23] are histograms examining the distribution of distances separating a solute atom from its nearest solute neighbor. The experimental K nearest neighbor distribution is presented in Fig. 2b in comparison to the distribution expected if K was distributed randomly in the system. There is a clear shift of the experimental distribution to the left relative to the random distribution. This indicates that there are significantly more than expected K atoms separated by small distances in the APT pure    single-A    single-B    pair    trio reconstruction. There is therefore a strong evidence to support the suggestion that K atoms in the 122 pnictide superconductor are not randomly distributed with possibility of the presence of K high and low solute regions at atomic scale.

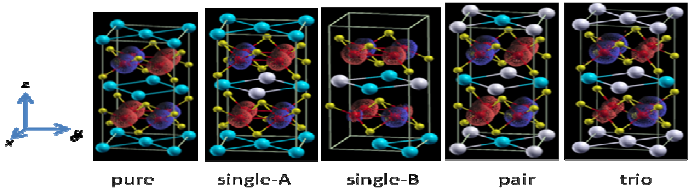

Fig.4 Isosurface plots of the spin density of the pure and various K-doped BaFe₂As₂ based on LDA. Large blue and white spheres indicate Ba and K atoms, and small red and yellow Fe and As, respectively. Note the nearly completely magnetism suppression at a fraction of the Fe atoms in the single-B and trio structures.

The localized K content fluctuation in $BaFe_2As_2$ matrix will lead to inhomogeneous superconducting transition and $T_c$ fluctuation, which can be evaluated by $T_c$ distribution. As proposed by Wang et al. [24], overall $T_c$ distribution across samples can be estimated from the specific heat measurements by means of a particular deconvolution method: $F(T) = \int_0^T f(T_c) dT_c$, where $F(T)$ is the integral of the distribution function, $f(T_c)$, and $F(T)$ represents the fraction in the sample with $T_c \leq T$. Fig. 3 shows the $F(T)$ and $f(T_c)$ dependences on temperature for $(Ba_{0.72}K_{0.28})Fe_2As_2$ crystals. The superconducting transition starts at 32.5 K and reaches maximum value of 31 K followed by a slow decrease down to 17 K. The broad transition of 15 K evidenced the influence of localized fluctuation in K content on the superconducting phases. Recent heat capacity measurements also suggest the presence of nanoscale inhomogeneity in the Co-doped $BaFe_2As_2$ system[25].

To understand of K doping effect, particularly the evolution of superconducting and magnetic properties with regard to the variation of K dopant concentrations observed above by APT, we perform extensive density-functional theory (DFT) calculations using local-density approximation (LDA) [26] for the exchange-correlation functional as implemented in the DMol³ code [27]. Double numerical quality localized basis set with a real-space cut off of 11 bohr and Monkhorst-Pack grids of 8×8×4 with 75 **k** points in the irreducible part of the Brillouin-zone for the 20-atom cell are employed. Polarization functions and scalar-relativistic corrections are incorporated explicitly. For the doped systems, we deliberately enforce no symmetry constraints. The low-temperature orthorhombic $F_{mmm}$ structure containing 0, 1, 2 and 3 K atoms substituting at Ba site, respectively, was employed to simulate the different solute regions. Full relaxation, including the atomic

Table I: Lattice constants, atomic magnetic moments and volume of undoped and K-underdoped $BaFe_2As_2$ obtained from DFT-LDA.

| Configuration | a, b, c (Å) | Magnetic Moment (μ_B) |
|---|---|---|
| Pure | 5.556, 5.503, 12.039 | ±0.47 |
| Pure (exp) [30] | 5.616, 5.571, 12.943 | ±0.47 [30] or ±0.99 [29] |
| Single-A | 5.461, 5.440, 12.171 | ±0.344 |
| Single-B | 5.455, 5.454, 12.285 | ±0.468, ±0.012, ±0.449, ±0.267 |
| Pair | 5.398, 5.376, 12.568 | 0.708, 0.706 |
| Trio | 5.345, 5.239, 12.703 | 0.065, 0.641 |

positions and lattice constants, is performed in all calculations.

The optimized structural data as well as the Fe-atomic magnetic moment are compiled in Table 1. The magnetic structures of undoped and underdoped systems are shown in Fig. 4. In agreement with experiment [10], our first-principles calculations reveals that replacing $Ba^{2+}$ with $K^+$ results in the continuous increase in $c$, but decrease in $a$ and $b$. For the pure $BaFe_2As_2$, the magnetic structure is ferromagnetic along $b$ but antiferromagnetic along $a$ and $c$ [28], with the calculated moment being 0.47 $μ_B$, compared with the experimental values of 0.99 $μ_B$ [29] and 0.87 $μ_B$ [30]. Such an underestimation is mainly due to LDA. For instance, using experimental lattice constants, LDA predicts the Fe moment being 0.68 $μ_B$, while the generalized gradient approximation being 1.98 $μ_B$. For the underdoped systems, importantly, we found that there is finite magnetism in all doping concentrations under LDA. The magnetic properties are extremely sensitive to local symmetry or doping environment. For the Trio and distorted Single-B (energetically degenerate to Single-A), atomic moment suppression occurs on only part of Fe atoms. Note in real doped sample, atomic distortion is always expected. While the underlining mechanism of the competition/coexistence model is not well understood at atomic level, $^{57}$Fe-Mössbauer measurements indicate a microscopic coexistence of the SDW and superconducting states [12]. Considering the fact that LDA-DFT general underestimates magnetism, our results suggest that the completely magnetism suppression unlikely occur in mesoscopic region; rather, the AFM-SDW and SC are expected to coexist in an even much more intimate level. Such a scenario is strongly supported by recent NMR investigation showing that these two states coexist microscopically in underdoped compositions, at least on a lattice parameter length scale [31]. Atomic coexistence of superconductivity and incommensurate magnetic order has been proposed theoretically [32], and been demonstrated in Co-doped $BaFe_2As_2$ [33].

Many superconductor studies make the assumption that the introduced carriers/dopants are distributed uniformly, leading to an electronically homogeneous in the system. However, the present work challenges this assumption, verifying the clustering of K dopant atoms. Substitution of K atoms to the Ba sites has two major influences in the 122 system: suppression of the Fe magnetism in its vicinity and providing hole carriers to promote superconductivity. The K inhomogeneity manifests as spatial variations in both the local density of states spectrum and the superconducting energy gap where it separates the electron systems into hole-rich and hole poor regions. This resembles the nanoscale superconducting gap variations observed in [13, 34] where Massee et al.[13] correlated nanometer-sized regions of different gap magnitude to the Co–Co interatomic separation. According to the K doping phase diagram [2], AFM-SDW is absent in the high K-solute regions (overdoped), creating an exclusively SC state, whereas coexistence of AFM and SC state appears in the low K (under-doped) regions and regions with no K are represented by a solely AFM state. Since the K clusters observed experimentally via APT in this study are only of the order of a few nanometer size, our results better support the picture of mixed phase separation (SC and AFM state) and the coexistence of AFM and SC occurred in lattice parameter length scale depending on the K concentration. The present observation is in agreement with the ref. [9, 10, 14, 35] that disorder in the K doped 122 compound could originate from inhomogeneity of K concentration. Such a situation resembles the widely discussed high $T_c$ cuprate superconductor, $Bi_2Sr_2CaCu_2O_{8+\delta}$, where the oxygen distributions are strongly correlated to nanoscale electronic disorder [36, 37]. Electronic phase segregation is linked to the formation of a clustering of the excess oxygen ions in the $La_2CuO_{4+y}$ system, [38]. Similar phenomena are likely to be common in all the pnictides system where randomness of dopant atom distribution plays an important role on the local electronic disorder.

In summary, by combination of APT, $T_c$ variations measurements and DFT simulations, we have addressed the underlying cause of nanoscale electronic state variations and coexistence of AFM and SC issues in K doped superconductors. APT results have provided direct experimental evidence that K atoms are non-uniformly distributed in $(Ba_{1-x}K_x)Fe_2As_2$ pnictide evidenced by the broad $T_c$ variation. DFT calculation indicates that AFM and SC may coexist on a lattice parameter length scale. The competition of AFM and SC is highly sensitive towards structural variation. Variation of K dopants in atomic scale results in a mixed scenario of phase coexistence and phase separation despites the strong local electronic inhomogeneities.

This work is supported by Australian Research Council Discovery Project (DP0984402) and AMMRF at the University of Sydney. WKY acknowledges L. Stephenson and A. Ceguerra for useful discussions.

Electronic address: waikong.yeoh@sydney.edu.au; rongkun.zheng@sydney.edu.au

[1] J. Zhao et al., Nat. Mater. 7, 953 (2008).

[2] H. Chen et al., Epl 85, 17006 (2009).

[3] H. Luetkens et al., Nat Mater 8, 305 (2009).

[4] J. T. Park et al., Phys. Rev. Lett. 102, 117006 (2009).

[5] T. Goko et al., Phys. Rev. B 80, 024508 (2009).

[6] H. Fukazawa et al., J. Phys. Soc. Jpn. 78, 033704 (2009).

[7] D. S. Inosov et al., Phys. Rev. B 79, 224503 (2009).

[8] A. A. Aczel et al., Phys. Rev. B 78, 214503 (2008).

[9] M. H. Julien et al., Epl 87, 37001 (2009).

[10] M. Rotter et al., New J. Phys. 11, 025014 (2009).

[11] P. Marsik et al., Phys. Rev. Lett. 105, 057001 (2010).

[12] F. L. Ning et al., J. Phys. Soc. Jpn. 78, 013711 (2009).

[13] F. Massee et al., Phys. Rev. B 79, 220517 (2009).

[14] D. Johrendt et al., Physica C 469, 332 (2009).

[15] T. F. Kelly et al., Rev. Sci. Instrum. 78, 031101 (2007).

[16] G. L. Sun et al., arXiv:0901.2728v3 [cond-mat.supr-con] (unpublished)

[17] X. L. Wang et al., Phys. Rev. B 82, 024525 (2010).

[18] M. Rotter et al., Angew. Chem.-Int. Edit. 47, 7949 (2008).

[19] D. Sanchez et al., Physica C:200, 1 (1992).

[20] P. Felfer et al., Ultramicroscopy In Press, , (2010).

[21] B. Gault et al., Ultramicroscopy in press, (2010).

[22] M. P. Moody et al., Ultramicroscopy 109 815 (2009).

[23] L. T. Stephenson et al., Microsc. Microanal. 13, 448 (2007).

[24] Y. Wang et al., Supercond. Sci. Technol. 19, 263 (2006).

[25] K. Gofryk et al., Phys. Rev. B 81, 184518 (2010).

[26] J. P. Perdew et al., Phys. Rev. B 45, 13244 (1992).

[27] B. Delley, The J. Chem. Phys. 92, 508 (1990).

[28] Akt et al., Phys. Rev. B 79, 184523 (2009).

[29] Y. Su et al., Phys. Rev. B 79, 064504 (2009).

[30] Q. Huang et al., Phys. Rev. Lett. 101, 257003 (2008).

[31] R. R. Urbano et al., Phys. Rev. Lett. 105, 107001 (2010).

[32] A. B. Vorontsov et al., Phys. Rev. B 79, 060508 (2009).

[33] Y. Laplace et al., Phys. Rev. B 80, 140501 (2009).

[34] Y. Yin et al., Phys. Rev. Lett. 102, 097002 (2009).

[35] Y. Laplace et al., Eur. Phys. J. B 73, 161 (2010).

[36] S. H. Pan et al., Nature 413, 282 (2001).

[37] K. McElroy et al., Science 309, 1048 (2005).

[38] H. E. Mohottala et al., Nat Mater 5, 377 (2006).